\documentstyle[epsf,prb,aps]{revtex}
\begin{document}
\twocolumn[\hsize\textwidth\columnwidth\hsize\csname @twocolumnfalse\endcsname
\draft
\title{Monte Carlo Simulation of Liquid-Crystal Alignment and of Chiral
Symmetry-Breaking}
\author{Jianling Xu and Robin L. B. Selinger}
\address{Physics Department,
Catholic University of America,
Washington, DC 20064}
\author{Jonathan V. Selinger and R. Shashidhar}
\address{Center for Bio/Molecular Science and Engineering,
Naval Research Laboratory, Code 6900, \\
4555 Overlook Avenue, SW,
Washington, DC  20375}
\date{June 11, 2001}
\maketitle

\begin{abstract}
We carry out Monte Carlo simulations to investigate the effect of molecular
shape on liquid-crystal order.  In our approach, each model mesogen consists of
several soft spheres bonded rigidly together.  The arrangement of the spheres
may be straight (to represent uniaxial molecules), Z-shaped (for biaxial
molecules), or banana-shaped (for bent-core molecules).  Using this approach,
we investigate the alignment of the nematic phase by substrates decorated with
parallel ridges.  We compare results for wide and narrow ridge spacing and
examine local order near the substrates, and show that our results are
consistent with the predictions of Landau theory.  We also investigate chiral
symmetry-breaking in systems of bent-core molecules.  We find a chiral
crystalline phase as well as a nonchiral smectic-A phase, but not a chiral
smectic-C phase.
\end{abstract}


\vskip2pc]
\narrowtext

\section{Introduction}

Molecular simulations have emerged as an important tool for investigating
liquid crystals.  The goal of the simulations is to relate the microscopic
structure of molecules to the macroscopic properties of liquid-crystal phases.
To do this, researchers begin with some assumptions about molecular structure,
simulate a large number of interacting molecules, and determine the large-scale
order of the resulting system as a function of temperature, applied fields, and
other variables.  This technique provides insight into the origin of
liquid-crystal phase transitions and textures.  It also assists in the
development of new liquid-crystal systems for technological applications.

Researchers have used several approaches for molecular simulations of liquid
crystals.  On the most microscopic scale, some studies begin with the structure
of liquid-crystal molecules in atomistic detail, and let the molecules interact
through a potential derived from the sum of interatomic
interactions.~\cite{glaser,glaser99}
The advantage of this approach is that it can
simulate actual experimental materials and can show how macroscopic properties
change with slight variations in the chemical structure.  The disadvantage is
that this approach is computationally intensive, and hence it can only be done
for relatively small systems, which may not be large enough to show all the
important features of liquid-crystal ordering.  Accuracy of classical
potentials can also be an issue.

Because of these limitations, many other studies begin with greatly simplified
models of molecular structures and interactions, which require less computer
time and hence allow larger system sizes.  Some studies represent molecules as
cylinders or spherocylinders, which interact through a hard or soft
excluded-volume repulsion.~\cite{photinos,frenkel,allen}  Other studies
represent molecules as ellipsoids interacting through the Gay-Berne potential,
an anisotropic generalization of the Lennard-Jones
potential.~\cite{gay-berne,luckhurst,zannoni}  A related approach uses the
Corner potential.~\cite{zewdie}  To model more complicated molecular
structures, further studies represent molecules as composites of two Gay-Berne
ellipsoids.~\cite{memmer}  Soft-sphere potentials can also be joined together
in a linear chain to represent molecules.  Such molecules can be
rigid~\cite{paolini} or semiflexible.~\cite{affouard,lee}

As an alternative approach, we model liquid-crystal molecules as a composite of
soft spheres bonded rigidly into a fixed shape, with no
intramolecular flexibility.  This approach is particularly useful for two
reasons.  First, the intermolecular interaction can be calculated at low
computational cost, thus allowing the simulation of large systems.  Second, the
spheres may be bonded into any shape to represent molecular structures of any
symmetry.  These features allow us to explore the complex relationship between
molecular shape and phase behavior, without undertaking full-scale atomistic
simulations.  The approach thus represents a compromise between the simplest
and most complex models used in liquid crystal simulation.

In a recent paper,~\cite{xu} we used this approach to simulate smectic liquid
crystals.  Our goal was to model the electroclinic effect, in which an applied
electric field induces molecular tilt in the smectic-A phase.  For that reason,
we considered model molecules with the structure shown in Fig.~1(a).  The
overall shape is a biaxial zig-zag shape like the letter Z, and a transverse
electric dipole moment makes the molecule chiral.  This shape was motivated by
the fact that many real liquid-crystal molecules exhibiting the electroclinic
effect have this general structure.  Indeed, the biaxiality of this structure
should be important for the electroclinic effect because a tilted smectic phase
is biaxial.  Our simulation results confirmed that the degree of biaxiality or
obliqueness in molecular shape is crucial for spontaneous molecular tilt in the
smectic-C phase and for induced molecular tilt in the smectic-A phase.

The purpose of our current study is to explore this simulation method further
by applying it to two different problems in the physics of liquid crystals.
First, we consider the alignment of a nematic liquid crystal by a ridged
surface.  Second, we explore chiral symmetry-breaking in the smectic phases of
a bent-core (or banana-shaped) liquid crystal.  In each case, we construct
model mesogens consisting of soft spheres bonded rigidly together with an
appropriate shape.  Our results for these two problems shows that this
simulation method is a versatile approach for modeling different types of
liquid-crystal ordering.

The first problem that we consider, the surface alignment of a nematic phase,
is a classic problem in liquid-crystal science and is fundamental to all
applications of liquid crystals.~\cite{blinov-chigrinov,tomilin}
Liquid-crystal cells are constructed so that the surfaces favor alignment of
the molecules in a particular direction, which may be planar (in a particular
direction within the surface plane), homeotropic (normal to the surface), or
between these extremes.  An applied electric field can cause the molecules in
the interior of the cell to tilt away from the alignment favored by the
surfaces.  Thus, the competition between surface alignment and an applied field
allows the molecules in the cell to be switched between different orientations,
changing the optical properties of the cell.  Key questions in designing new
surfaces to optimize liquid-crystal alignment are:  What type of surface is
needed to align liquid-crystal molecules?  How far into a cell does the
alignment persist?

To address these questions, we perform Monte Carlo simulations of nematic
alignment, using the simulation approach discussed above.  For the model
molecules, we would like to construct a rigid uniaxial rod-like structure,
which is known to give a nematic phase at high enough densities.  For that
reason, we use seven spheres bonded together in a straight line, as shown in
Fig.~1(b).  For the aligning surface, we would like to model a rubbed polymer
substrate, which has parallel ridges that give planar alignment of the liquid
crystal.  We use a flat surface decorated with regularly spaced elevated
ridges, each composed of spheres fixed in a row, as shown in Fig.~2.  Our
simulation results show that the nematic director aligns with the ridges for
all ridge spacings studied.  The nematic order parameter near the substrate
increases as the ridge spacing is reduced, indicating that narrowly spaced
ridges on the substrate suppress director fluctuations.  However, the
correlation length associated with the drop-off of nematic order away from the
substrate is independent of ridge spacing, and is a characteristic of the bulk
nematic phase.

Our second problem, chiral symmetry-breaking in systems of bent-core molecules,
concerns a new type of smectic liquid crystal.  In recent years, researchers
have synthesized liquid-crystal molecules with a bent or ``banana-shaped''
core.~\cite{niori,link,heppke}  The individual molecules are not chiral.  In
the bulk or in freely suspended films, the molecules form smectic phases.  In
the smectic-A phase, the molecules are on average normal to the smectic layers,
and hence the phase is not chiral.  However, when the material is cooled into a
smectic-C phase, the molecules can tilt with respect to the layers in two
inequivalent ways that are mirror images of each other.  For that reason, the
system undergoes a chiral symmetry-breaking transition, in which it forms
right- and left-handed chiral domains.~\cite{niori,link,heppke}
There are actually several different
smectic-C phases, depending on how the domains of alternate chirality and
alternate tilt pack together.  Key questions for the theory of these systems
are:  What type of intermolecular interaction drives the chiral
symmetry-breaking transition?  How do the molecules pack together in the
achiral and the chiral phases?

To investigate these issues, we perform Monte Carlo simulations of bent-core
liquid crystals, again using the same simulation approach.  In this case, we
construct model molecules by arranging seven soft spheres in a rigid bent-core
shape, as shown in Fig.~1(c).  In the simulations, we find that a bulk system
of these molecules forms a nonchiral smectic-A phase at high temperatures, and
a chiral tilted crystalline phase at low temperatures.  The chiral
symmetry-breaking transition is driven by a favored packing of the molecules,
which can fit together more efficiently in either a right- or a left-handed
state than in a symmetric state.  Thus, at least in this model, the steric
interaction of rigid molecules is sufficient for chiral symmetry-breaking;
neither molecular flexibility nor dipole-dipole interactions are necessary.
Although this system forms a chiral tilted phase with three-dimensional
crystalline order, it does not form a chiral smectic-C phase with only
one-dimensional positional order.  We believe that the simultaneous onset of
chiral order and crystalline order is an artifact of our particular choice of
molecular shape, because the same type of chiral order can occur in either a
crystalline or a smectic-C phase.

\section{Surface Alignment of Nematic Phase}

To simulate a nematic phase, we use the rod-like model molecule shown in
Fig.~1(b). Each molecule is composed of seven spherical particles, or
interaction sites, fixed rigidly in a straight line, with no intramolecular
degrees of freedom. The sphere-sphere interaction potential is the truncated
Lennard-Jones potential cut off at its minimum so there is no attractive tail,
also known as the Weeks-Chandler-Andersen potential,~\cite{wca}
\begin{equation}
U^{\rm int}_{mn}=
\cases{\displaystyle 4\epsilon\left[
\left({\sigma\over r_{mn}}\right)^{12}-
\left({\sigma\over r_{mn}}\right)^{6}
\right]+\epsilon,
&if $r_{mn}\le r_{c}=2^{1/6}\sigma$;\cr
0,&otherwise.\cr}
\label{wcapotential}
\end{equation}
where $r_{mn}=|{\bf r}_m - {\bf r}_n|$ and $m$ and $n$ are the sphere indices
referring to different molecules.  In this way, the steric interactions between
molecules are represented via a short-range repulsive potential, with no
attractive or electrostatic interactions.

The simulation cell consists of 5,070 molecules in a three-dimensional
rectangular box of size $35 \times 35 \times 41$, where lengths are measured in
units of the Lennard-Jones parameter $\sigma$, and each molecule is seven units
long and one unit wide.  The system has periodic boundary conditions along $x$
and $y$, and is confined in the $z$ direction between parallel substrates.

Each substrate consists of a soft repulsive wall that interacts with the
spherical particles in each molecule using the same WCA potential.  Each
substrate is decorated with regularly spaced elevated ridges composed of
spheres fixed in parallel rows along the wall, as shown in Fig.~2.  Each of
these spheres is identical to the spheres making up the liquid-crystal
molecules, and again has the same WCA interactions.  Thus, interactions between
the liquid crystal and the wall are essentially steric in nature, with no
attractive or electrostatic terms.

Each wall has an integer number of ridges, spaced evenly.  We make the ridge
pattern the same for the top and the bottom substrates, so that the nematic is
not twisted, having the same preferred planar orientation on both substrates.
However we slightly shift the ridge patterns so that the substrates do not have
an exact mirror symmetry.

We perform Monte Carlo (MC) simulations at the fixed temperature $k_B T=5.0$,
well above the nematic--smectic-A transition point, which is about 3.5,
measured in units of the Lennard-Jones parameter $\epsilon$.  The density is
held fixed at $\rho=0.7$ spheres/unit volume.  In preliminary simulations, we
either randomize the initial molecular orientations or initialize them in a
homeotropic orientation, aligned normal to the substrates.  We find that either
of these initial conditions leads to a slow ordering process, which eventually
gives rise to an aligned planar structure, with the nematic director parallel
to the ridges on the substrates.  After determining the preferred orientation,
we perform further simulations with the molecules initialized in the preferred
planar orientation, with the molecules aligned parallel to the ridges and the
center-of-mass positions randomized.

Each MC step consists of three attempted translations (along $x$, $y$, and $z$)
and two attempted rotations (in $\theta$ and $\phi$) per molecule, with
appropriate sampling in the polar angle.~\cite{allen-tildesley}  We carry out
100,000 MC steps to reach equilibrium and then an additional 200,000 MC steps
to collect data.

To measure the spatial variation of liquid-crystal order near the
substrates, we calculate the nematic order tensor $Q_{\alpha\beta}$, which
represents the strength and direction of orientational
order,~\cite{chaikin-lubensky}
\begin{equation}
Q_{\alpha\beta}=\left\langle\frac{1}{N}\sum_{j=1}^{N}
\left(\frac{3}{2}n_{j\alpha}n_{j\beta}-\frac{1}{2}\delta_{\alpha\beta}\right)
\right\rangle,
\label{Q}
\end{equation}
where ${\bf n}_j$ is the director along the axis of molecule $j$ and $N=5070$
is the number of molecules.  For any $z$, the largest eigenvalue of
$Q_{\alpha\beta}(z)$ represents the scalar nematic order parameter $S(z)$.  The
corresponding eigenvector identifies the local direction of alignment.

We evaluate $Q_{\alpha\beta}$ as a function of distance from the
substrate by dividing the system into fifty layers and calculating
$Q_{\alpha\beta}$ in each layer.  Because the number of molecules in
each layer is relatively small, around 100, we must average over time to
get a statistically meaningful sample.  Accordingly, we average over a
series of 100 independent configurations separated in time by 2000
MC steps each.

We perform simulations using substrates with three different ridge spacings,
but with the same bulk density $\rho=0.7$ spheres/unit volume, the same
temperature $k_B T=5.0$, and the same number of molecules $N=5070$ (a total of
35,490 particles, plus those decorating the substrate).  The number of ridges
on the substrates is chosen as 1, 2, and 7, with spacing between ridges of 35,
17.5, and 5 length units, respectively (taking periodic boundary conditions
into account).  The box dimensions are set at $H_x=35$, $H_y=35$, and
$H_z\approx 41$, with $H_z$ adjusted slightly to make the bulk densities the
same for all three systems, as the total number of particles in the box is
slightly increased by the addition of extra ridges.

In each case we find that the nematic director {\it throughout the cell} aligns
parallel to the ridges.  Figure~3 shows the nematic order parameter $S(z)$ vs.\
$z$ for all three systems, with one substrate at $z=0$ and the other at
$z=H_z$.  The bulk order parameter is about the same in all cases, but the
order near the substrate varies with ridge spacing.  The narrowest ridge
spacing gives the largest enhancement in local order, while the widest ridge
spacing gives the least enhancement.  Figure~3 also shows that the correlation
length associated with nematic order near the interface is independent of the
ridge spacing.  In other words, no matter how densely spaced the substrate
ridges, the thickness of the ordered interface does not change.  We fit the
data in Fig.~3 to the form
\begin{equation}
S(z)=S_0+A (e^{-z/\xi}+e^{-(H_z - z)/\xi}),
\end{equation}
where $S_0$ is the bulk order parameter value, $A$ is the amplitude of
enhancement near the substrates, and $\xi$ is the associated correlation
length.  Using the least-squares algorithm, we fit the data for all three ridge
spacings.  Figure~3 shows an example of the fit for the ridge spacing of 5
units, and Table I gives the fit parameters.

Our results can be understood through an analytical calculation using a Landau
free energy functional, which is similar to the model in
Ref.~\onlinecite{sheng}.  We
assume that the nematic order parameter takes the form,
\begin{equation}
S(z)=S_{0}+S_{1}(z)
\end{equation}
where $S_0$ is the bulk value and $S_1(z)$ is the surface-induced contribution.
Near a single aligning substrate, the free energy can be written as
\begin{equation}
F=\int dz \left[\frac{1}{2}K\left(\frac{dS}{dz}\right)^2
+\frac{1}{2}a(S(z)-S_0)^2 \right] - hS(z=0).
\end{equation}
where $a$ and $K$ are bulk parameters dependent on temperature and density.
The term $-hS(z=0)$ favors enhanced nematic order at the substrate, and one
expects the anchoring strength $h$ to vary with substrate structure.

Minimizing the free energy functional gives
\begin{equation}
S(z)=S_0 +A e^{-z/\xi},
\end{equation}
where $A=h/\sqrt{K a}$ and $\xi=\sqrt{K/a}$.  These equations show that the
amplitude $A$ is a surface parameter, which depends on $h$ as well as $K$ and
$a$ and thus varies with substrate structure.  By contrast, the correlation
length $\xi$ is a bulk parameter, which depends only on $K$ and $a$ and thus
varies with temperature and density but is independent of substrate structure.
This is consistent with the simulation results for $A$ and $\xi$ in Table I.

Apart from the enhanced nematic order near the substrates, the simulations also
show a local tendency to {\it smectic\/} ordering near the substrates, even
though the temperature is well above the bulk nematic-smectic transition.
Figure~4 shows a snapshot of the molecular configuration near one substrate,
for the ridge spacing of 5 units.  Clearly the molecular density is not
uniform. Instead, there are local regions of incipient molecular layering in
the direction along the ridges.  This local layering suggests that the system
is near a surface nematic-smectic transition.  The tendency to smectic ordering
in a nematic close to a surface with planar orientation has also been studied
theoretically in the case of a bare substrate.~\cite{poniewierski}  In that
system, as in our system, the wall suppresses local director fluctuations,
giving rise to a more ordered state, just as an applied electric field can
drive a nematic-smectic transition.~\cite{xu}

In experiments on nematic alignment by buffed surfaces, it has sometimes been
observed that a nematic aligns in a direction different from, or even
perpendicular to, the buffing axis.~\cite{kobayashi}  While our simulations did
not show that kind of ordering, our results lead us to consider possible
explanations.  One is that the local smectic order next to the substrate could
form with the layers parallel to the ridges, and with the molecules oriented
perpendicular to the ridges or at some degree of tilt.  It is possible that a
wider ridge spacing, a different molecular shape, or a temperature closer to
the nematic-smectic transition could produce such effects in a simulation like
ours.  Another likely explanation is that electrostatic or attractive
interactions play an important role in determining the preferred orientation.
These effects have not been considered in the present model but could be
introduced in future work.

\section{Chiral Symmetry-Breaking in Systems of Bent-Core Molecules}

As discussed in the Introduction, recent experimental studies have found that
liquid crystals composed of certain achiral molecules can undergo a spontaneous
symmetry-breaking and form chiral domains.~\cite{niori,link,heppke}  These
``bent-core'' or ``banana-shaped'' molecules form layered structures similar to
those observed in ferroelectric smectic liquid crystals, with the molecular
axis tilted with respect to the layer normal.  Each molecule has a net dipole
moment which defines a polar axis perpendicular to the long axis of the
molecule.  The chirality of the tilted phase is determined by the spatial
relationship of the layer normal, the tilt direction, and the polar axis; these
can be arranged in either right- or left-handed fashion.  The two arrangements
are degenerate in energy, leading to the formation of right- and left-handed
domains with equal probability.

We perform Monte Carlo simulations to investigate what type of intermolecular
interaction can drive the chiral symmetry-breaking transition, and how the
molecules pack together in the achiral and chiral phases.  In these
simulations, we use the same approach discussed above for simulations of
nematic alignment, but we change the model molecular structure in order to
represent bent-core molecules.  The model structure consists of seven spheres
arranged in a rigid bent-core shape, like the letter V, as shown in Fig.~1(c).
The bend angle between the two straight segments of the molecule is
$140^\circ$.  As in the earlier simulations, the molecules interact through the
soft repulsive sphere-sphere interaction of Eq.~(\ref{wcapotential}).  We
neglect dipole-dipole interactions in order to see what ordering is induced by
purely steric interactions.

To characterize the magnitude and sign of chirality of the system, we introduce
a chiral order parameter $\chi_i$ for each molecule $i$.  In a smectic phase,
let the unit vector ${\bf\hat{m}}_i$ represent the local layer normal,
${\bf\hat{n}}_i$ represent the long axis of the molecule, and ${\bf\hat{p}}_i$
represent the molecular dipole moment.  The chiral order parameter can then be
defined as
\begin{equation}
\chi_i = [({\bf\hat{m}}_i \times {\bf\hat{n}}_i) \cdot {\bf\hat{p}}_i]
[{\bf\hat{m}}_i \cdot {\bf\hat{n}}_i] .
\label{orderparameter}
\end{equation}
Note that this order parameter is invariant under the two symmetry operations
${\bf\hat{m}}_i\to-{\bf\hat{m}}_i$ and ${\bf\hat{n}}_i\to-{\bf\hat{n}}_i$, but
{\it not} under ${\bf\hat{p}}_i\to-{\bf\hat{p}}_i$, which is not a symmetry of
the system.  The average chiral order parameter is then
\begin{equation}
\bar{\chi}=\frac{1}{N}\sum_{i=1}^N \chi_i .
\end{equation}
That average order parameter can be compared with the rms chiral fluctuations
\begin{equation}
\chi_{\rm rms}=
\left[{\frac{1}{N}\sum_{i=1}^{N}(\chi_i-\bar{\chi})^2}\right]^{1/2}.
\end{equation}

We simulate 600 molecules in a rectangular box with periodic boundary
conditions.  We choose the height of the box so that the molecules can form
four smectic-like layers.  The density is held fixed at $\rho=0.7$ spheres per
unit volume.  At the temperature ${k_B}T=5.0$, the systems forms a {\it chiral
crystalline\/} phase.  A snapshot of this phase is shown in Fig. 5.  In
equilibrium, the average chiral order parameter is $\bar{\chi}=0.24$, while the
rms average chiral fluctuations are $\chi_{\rm rms}=0.06$.  The average chiral
order parameter is much greater than the chiral fluctuations, which means the
system is in a well-defined chiral state.  This spontaneous chiral ordering can
be seen in the snapshot, which shows the molecules tilting in a way that breaks
reflection symmetry.

The system shown in Fig.~5 has a large spontaneous tilt in each layer, but the
direction of the tilt appears to vary randomly from layer to layer.  The
coupling between adjacent layers due to excluded volume interactions is
apparently too weak to order the tilt directions.  A similar variation in the
tilt from layer to layer was found in the smectic-C phase of similar
molecules.~\cite{xu}  Although the tilt directions vary, the chiral order
parameter has the same sign in each layer, indicating that the simulation cell
represents a domain of uniform chirality.

The structure of a single layer, shown in Fig. 5(b), shows that the system has
periodic order with some disordered regions, so it is a crystalline rather than
a smectic phase.  When we raise the temperature or lower the density of this
system, the tilt vanishes and we find a stable untilted smectic-A phase.  We do
not observe any tilted chiral smectic-C phase.  In this system, the onset of
chiral order apparently occurs together with the onset of crystalline order.

The observation of a chiral phase, even though it is crystalline rather than
smectic, answers the basic questions raised above.  First, the simulations show
that the steric interaction of rigid molecules is sufficient to drive the
chiral symmetry-breaking transition.  Because we neglect dipole-dipole
interactions in the simulations, we see that they are not necessary for this
transition.  Furthermore, because we completely suppress intramolecular degrees
of freedom, we see that chiral ordering can occur in achiral materials without
the involvement of chiral excited molecular states.  Second, the snapshot shows
how the molecules pack together in the chiral phase.  We see that the molecules
tilt so that the spheres of one molecule fit into the spaces between the
spheres of adjacent molecules.  Thus, a tilted chiral state with either
handedness is a more efficient packing of neighboring molecules than an
untilted achiral state.  This shows an explicit mechanism for chiral
symmetry-breaking in a system with purely steric interactions.

In fact, the ``bumpy'' nature of molecular shape may play a crucial role in
favoring the tilted geometry, via the details of molecular packing.  In our
model, two parallel molecules do not pack together efficiently unless they are
tilted enough to allow the close packing of the component spheres.  A bent-core
molecule composed of two spherocylinders has also been simulated,~\cite{lansac}
but it does not show a chiral symmetry-breaking transition, presumably because
close packing of spherocylinders does not require molecular tilt.

We do not think that there is any fundamental connection between the onset of
chiral order and the onset of crystalline order.  The same type of chiral order
described by the order parameter of Eq.~(\ref{orderparameter}) can occur in
either a crystalline or a smectic-C phase.  The fact that we do not observe a
chiral smectic-C phase is probably an artifact of the particular molecular
structure that we have simulated.  To test how sensitively the phase behavior
depends on molecular structure, future simulations should make slight
variations on the molecular structure.  In general, a less sharp bend-angle
will favor a smectic phase over a crystalline phase at a given temperature and
density.  A bump strategically placed at the center of the bent-core molecule
will change the packing geometry and might favor twist (of either handedness)
over parallel packing.  Such a bump can be formed by increasing the interaction
distance of the central sphere, or by adding an extra sphere to the molecule.
The inclusion of dipole-dipole interactions could also destabilize the crystal
and favor the formation of a smectic phase.

In summary, we have extended a Monte Carlo simulation method, originally
developed for simulations of the electroclinic effect, to model surface
alignment of nematic liquid crystals and chiral symmetry-breaking in bent-core
liquid crystals.  The results give insight into the local molecular
arrangements that lead to both of these macroscopic effects.  These two studies
show that this simulation method is a versatile technique for investigating a
wide range of ordering phenomena in liquid crystals.

\acknowledgments

This research was supported by the Office of Naval Research and the Naval
Research Laboratory.

\epsfclipon

\begin{figure}
\centering\leavevmode\epsfxsize=3.375in\epsfbox{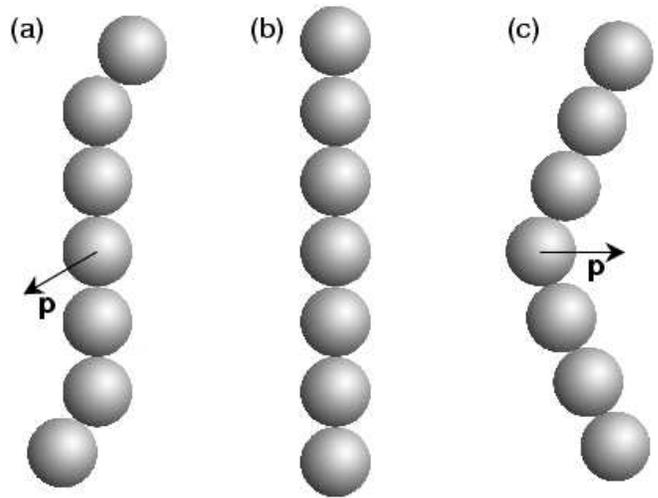}\bigskip
\caption{Structure of the model molecules discussed in this paper:  (a)~Biaxial
Z-shaped molecules, used in simulations of the electroclinic
effect.~\protect\cite{xu}  The transverse dipole moment $\bf p$ makes these
molecules chiral.  (b)~Uniaxial rod-like molecules, used in simulations of
surface alignment of the nematic phase. (c)~Bent-core (banana-shaped)
molecules, used in simulations of chiral symmetry-breaking.  Because the dipole
moment $\bf p$ is in the molecular plane, the molecules are not chiral.}
\end{figure}

\begin{figure}
\centering\leavevmode\epsfxsize=3.375in\epsfbox{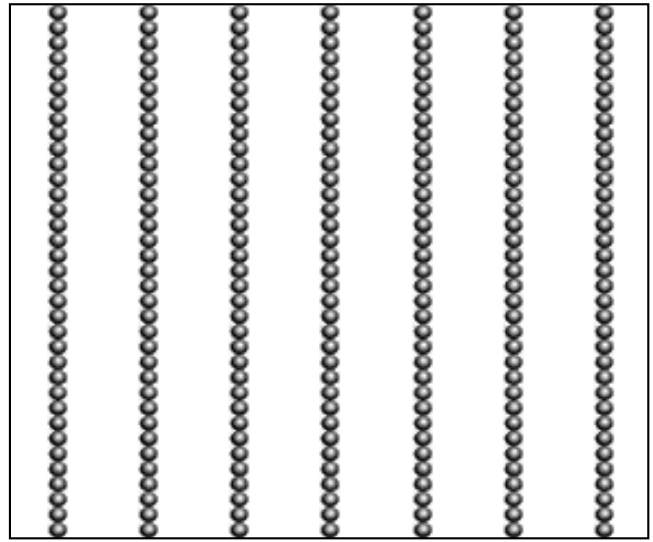}\bigskip
\caption{Flat substrate decorated with regularly spaced elevated ridges, each
composed of spheres fixed in a row.  In this substrate, the ridge spacing is 5
(in units of the Lennard-Jones diameter of the spheres).}
\end{figure}

\begin{figure}
\centering\leavevmode\epsfxsize=3.375in\epsfbox{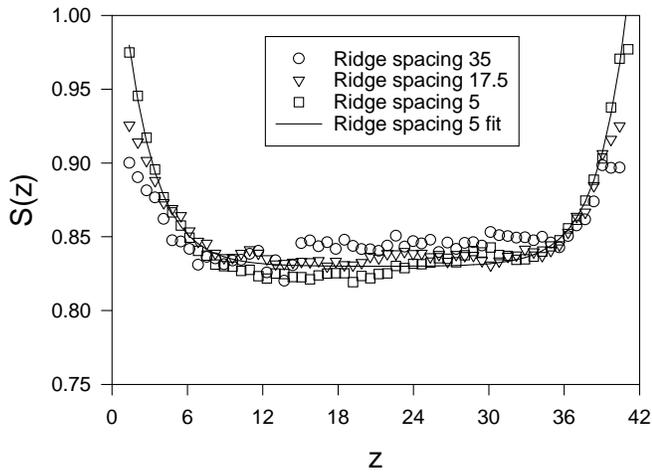}\bigskip
\caption{Profile of the nematic order parameter $S(z)$ as a function of $z$,
the distance across the thickness of the cell, for simulation runs with three
different ridge spacings on the cell surfaces.  Note the enhanced order near
the surfaces.}
\end{figure}

\begin{figure}
\centering\leavevmode\epsfxsize=3.375in\epsfbox{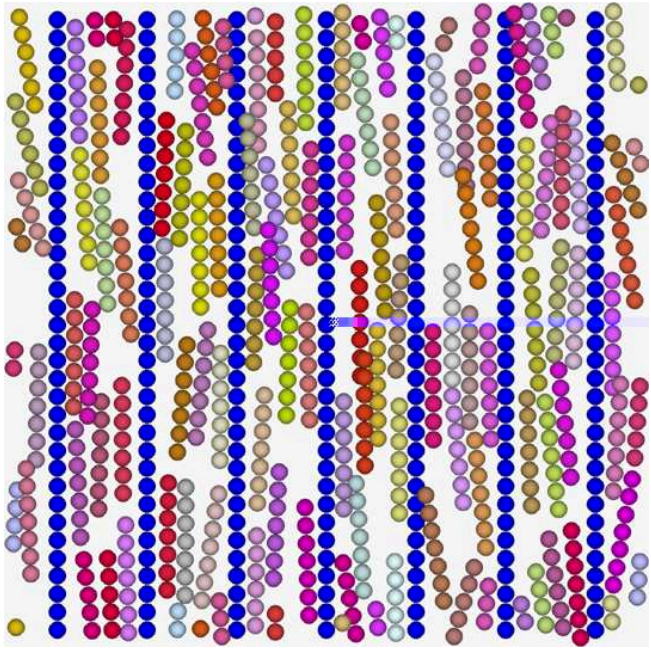}\bigskip
\caption{Snapshot of the molecular configuration near a substrate (with ridge
spacing of 5 units).  The long vertical rows of spheres are the ridges, while
the seven-sphere rods are the molecules.  Although the bulk phase is nematic,
the region near the substrate shows a local tendency to smectic ordering in the
direction along the ridges.}
\end{figure}

\begin{figure}
\centering\leavevmode(a)\epsfxsize=3.2in\epsfbox{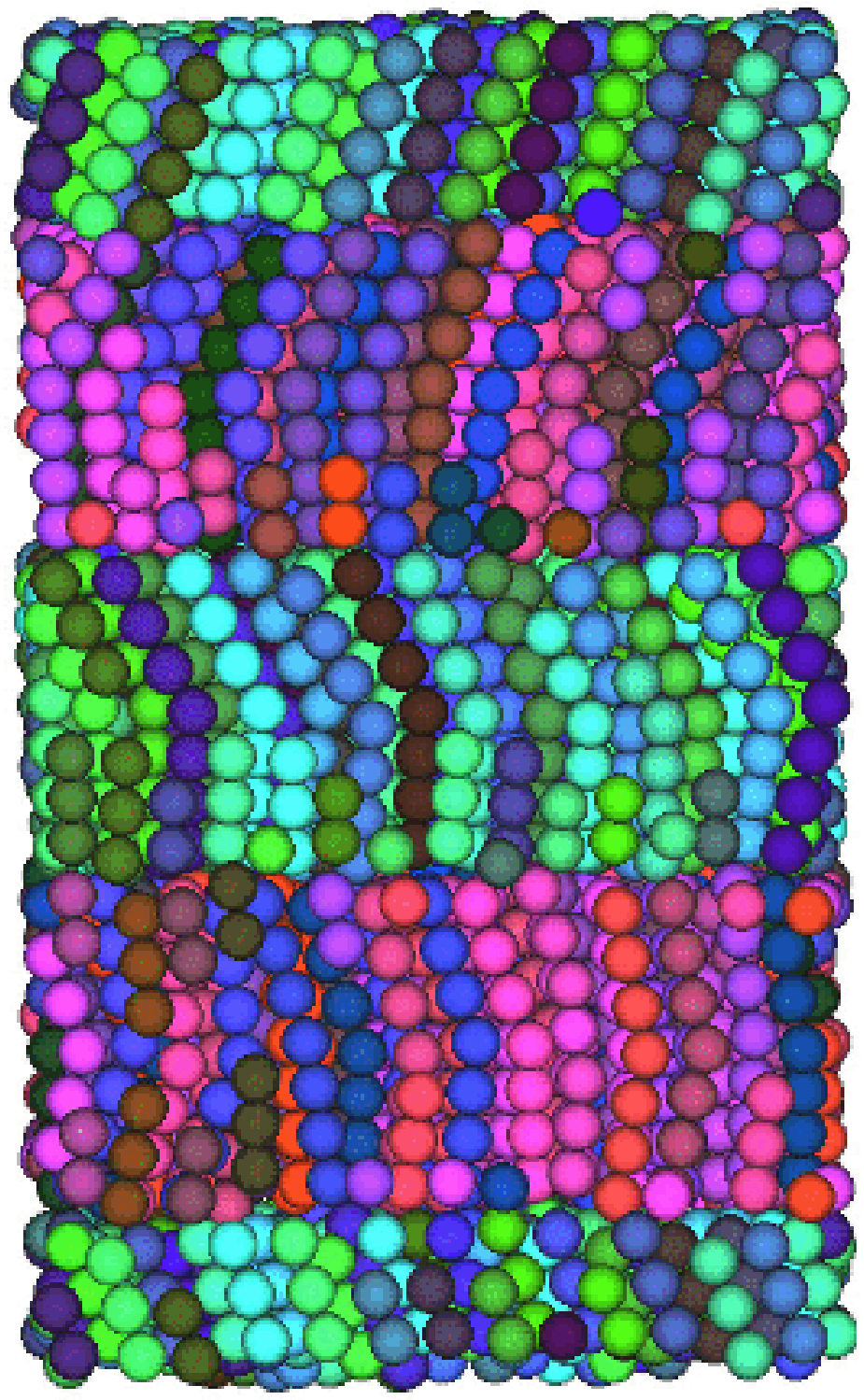}\bigskip

\centering\leavevmode(b)\epsfxsize=3.2in\epsfbox{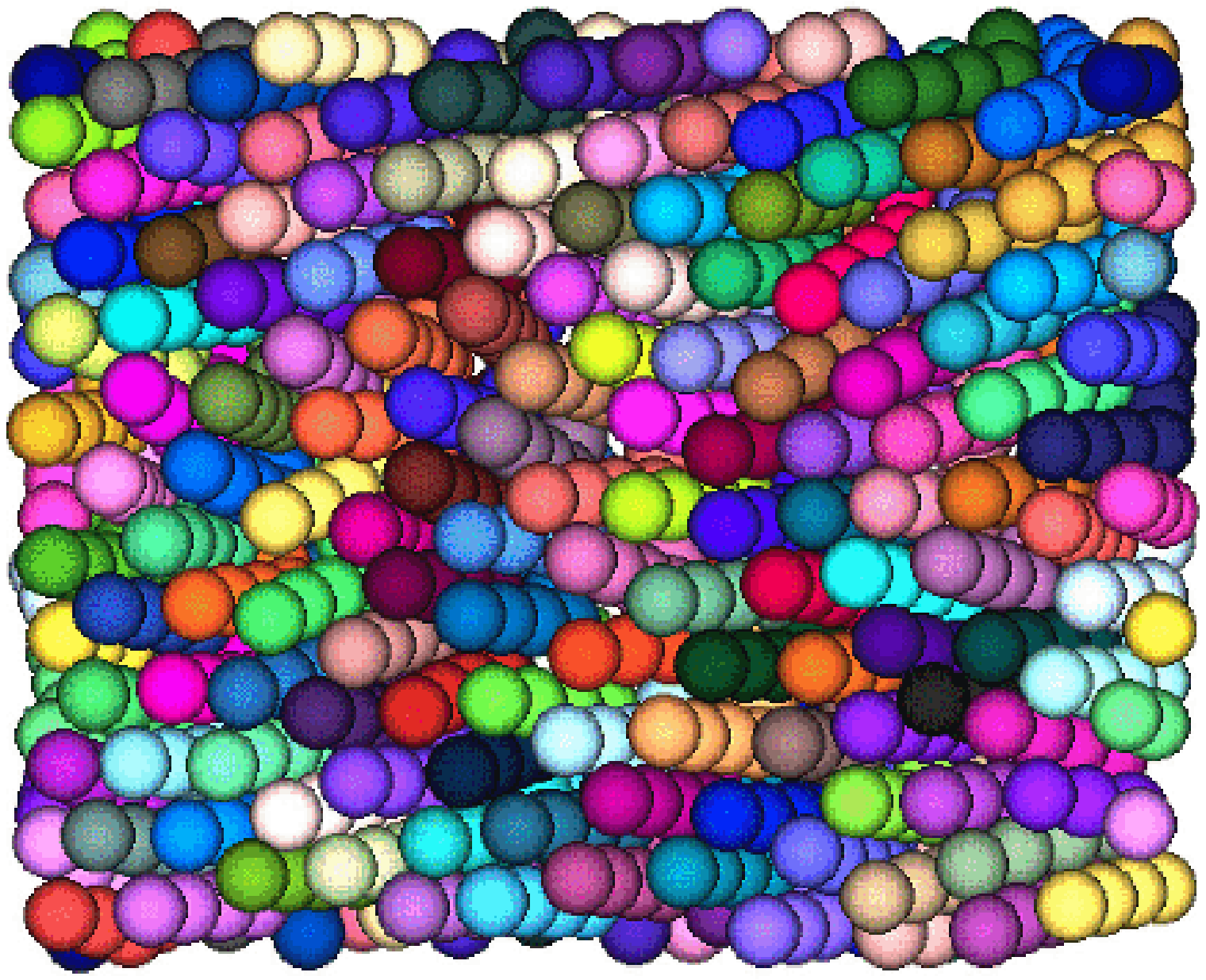}\bigskip
\caption{Snapshot of the configuration of bent-core molecules in a chiral
crystalline phase.  (a)~Side view.  (b)~Top view of a single layer.}
\end{figure}

\begin{table}
\caption{Fit parameters for the simulations of nematic alignment at three
different ridge spacings.}
\begin{tabular}{ddd}
Ridge spacing&$A$&$\xi$\\
\tableline
5.0&0.175&2.3\\
17.5&0.12&2.5\\
35.0&0.08&2.1\\
\end{tabular}
\end{table}

\end{document}